\documentstyle[twoside,fleqn,nacho,epsfig]{article}
\newcommand{\ptr}      {\mbox{$p_{\rm T}$}}
\newcommand{\ptds}      {\mbox{$p_{\rm T}(D^*)$}}
\newcommand{\etads}      {\mbox{$\eta (D^*)$}}

\newcommand{\dspm}       {\mbox{$ D^{*\pm}$}}

\newcommand{\ds}       {\mbox{$ D^{*}$}}

\newcommand{\pb}      {\mbox{pb$^{-1}$}}

\newcommand{\xd}      {\mbox{$x(D^*)$}}

\newcommand{\ftwoccb}   {\mbox{$F_2^{c\bar{c}}$}}
\newcommand{\ftwo}      {\mbox{$F_2$}}
\newcommand{\qsq}      {\mbox{$Q^2$}}
\newcommand{\y}        {\mbox{$y$}}

\newcommand{\mc}       {\mbox{$m_c$}}

\newcommand{\AmS}{{\protect\the\textfont2
  A\kern-.1667em\lower.5ex\hbox{M}\kern-.125emS}}

\title{Measurement of \dspm\ Cross Sections 
                   and the Charm Contribution to the 
Structure Function of the Proton 
                    in Deep Inelastic Scattering at HERA }

\author{I. Redondo \address{Dept. de F\'{\i}sica Te\'orica,
        Universidad Aut\'onoma de Madrid,
       Cantoblanco, Madrid 28049, SPAIN. e-mail: redondo@mail.desy.de}
\thanks{Supported by the Comunidad Aut\'onoma de Madrid.}.
On behalf of the ZEUS Collaboration}

\begin{document}

\begin{abstract}
We present measurements on \dspm\ production cross sections in deep inelastic 
 $e^+p$ scattering 
with $1<Q^2<600\,$GeV$^2$ and $0.02<y<0.7$ in two restricted kinematic regions in \ptr(\dspm) and $\eta$(\dspm).
The decay channels 
$D^{\ast +}\rightarrow D^0 \pi^+ $  with subsequent decay 
$D^0 \rightarrow K^- \pi^+$ or $D^0 \rightarrow K^- \pi^+ \pi^+ \pi^-$(+\,c.c.) are used.
The cross sections are extrapolated to the full
kinematic region to determine the charm contribution to the proton structure function.

\end{abstract}

\maketitle
\section{INTRODUCTION}
\vspace{-0.1cm}
The first HERA measurements of the charm contribution 
to the proton structure function, \ftwoccb\, were reported by 
the H1 and ZEUS Collaborations from an analysis of \dspm\
production in their 1994 data sets \cite{h194,z94}. The results 
were consistent with Photon Gluon Fusion (PGF) being
the dominant mechanism for \dspm\ production in $e^+p$ 
Deep Inelastic Scattering (DIS). If this is the case, this type 
of measurements are sensitive to the gluons in the 
proton. In addition, they can provide a test of the 
universality of the parton distribution functions (pdf's), 
namely, whether pdf's extracted from the inclusive measurement 
of the proton structure function, \ftwo, can be 
used as input for calculations of more exclusive processes as 
charm production.

Here we present a study 
of \dspm\ production using the 1996 and 1997 data corresponding to an 
integrated luminosity of 37 \pb. More than tenfold larger data sample,  
together with the modifications of the ZEUS detector made for the 1996 
and 1997 operation allow an extension of the kinematic range to both smaller 
and larger \qsq. The \dspm\ is tagged via  $D^{\ast +}\rightarrow (D^0 \rightarrow K^- \pi^+) \pi^+ $(+\,c.c.) and 
$D^{\ast +}\rightarrow (D^0 \rightarrow K^{-} \pi^+ \pi^+\pi^-) \pi^+ $(+\,c.c.) decay channels, refered to as  $K2\pi$ and $K4\pi$ respectively.
\section{\dspm\ CROSS SECTIONS}
\begin{figure*}[htb]
\vspace{-1.cm}
\epsfig{file=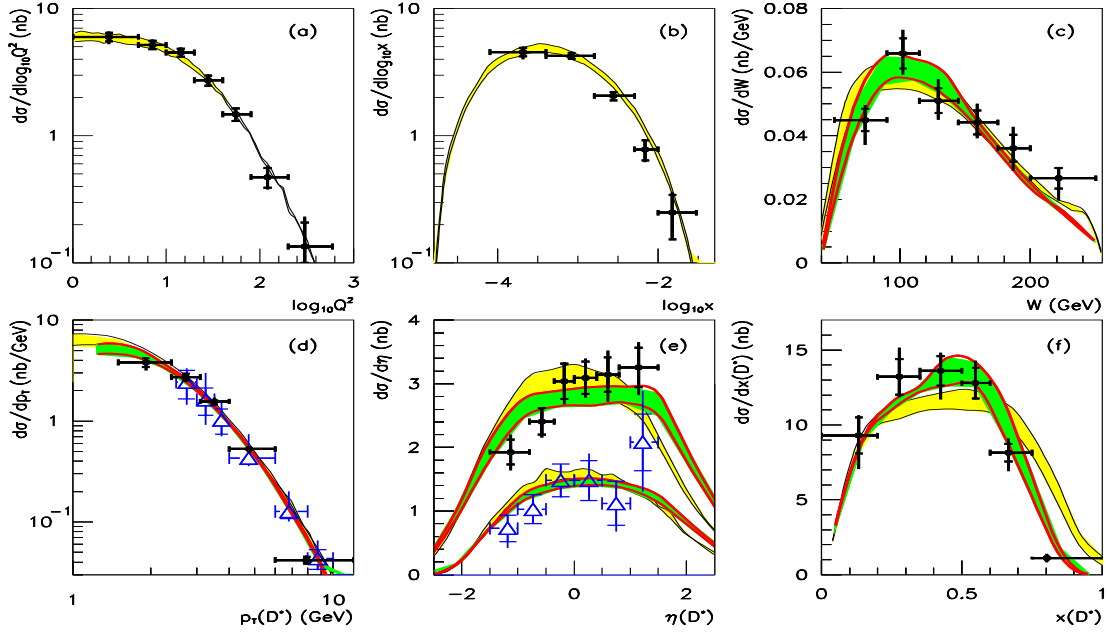,width=16cm,height=9.5cm}
\vspace{-1.5cm}
\caption{ Differential cross sections for \ds\ production from the $K2\pi$ final state (solid dots) and from the $K4\pi$ channel (open triangles).
The light shaded band corresponds to the
standard Peterson fragmentation function. The dark shaded band show the NLO reweighted RAPGAP MC. The two sets of bands correspond to charm mass 
variations between 1.3 (upper curve) and 1.5~GeV (lower curve). In (a) and (b)
the dark band is indistinguishable from the light one and is not shown.
}
\label{diffxsections}
\end{figure*}

\vspace{-0.1cm}
The measured \dspm\ cross section using the $K2\pi(K4\pi)$ final state, 
in the region $1.5(2.5)<\ptds<15 ~ GeV$, $|\etads|<1.5$ is
$\sigma$\small{$(e^{+}p\rightarrow e^{+} \dspm X)$} \normalsize{$= 8.31\pm 0.31(stat)^{+0.30}_{-0.50}(sys)$ nb}($3.65 \pm 0.36(stat)^{+0.20}_{-0.41}(sys)$ nb).
Figure~\ref{diffxsections} shows the differential \ds\ cross sections 
in the restricted \qsq, \y, \ptds\ and \etads\ region as functions of  $log_{10}(\qsq)$, $log_{10}(x)$, $W$, \ptds, \etads\ and \xd $=2p^*(\ds)/W$,  where $p^*(\ds)$ is the momentum in the $\gamma^*$-proton CMS frame.
The results using each decay channel can be directly 
compared in the \ptds\ differential cross 
section. The agreement is satisfactory. 

\section{COMPARISON WITH NLO QCD}
\vspace{-0.1cm}
We compute NLO QCD calculations with a semi-inclusive Monte Carlo generator HVQDIS\cite{hvqdis}
for heavy quark production and subsequent fragmentation to \dspm\ via a Peterson fragmentation function~\cite{peterson} .
This generator
is based on NLO calculations~\cite{smith} 
in the three flavor number scheme (TFNS),
in which 
only light quarks ($u,~d,~s$) are included in the initial state 
proton. Heavy quarks are produced exclusively
by the convolution of the light flavours and the gluon
with the massive matrix elements
and coefficient functions calculated previously~\cite{laenen}.

We use as 
input pdf ZEUS94~\cite{zeusfit}. The QCD renormalization and factorization scales are 
set to $\sqrt{\qsq + 4\mc^2}$. \mc\ is varied between 1.3 and 1.5 GeV. $f(c\rightarrow\ds)=0.222$ is taken from $e^+e^-$ measurements
~\cite{OPAL222}. The error on this quantity introduces a
normalization uncertainty of $\sim 9$\%.
 Finally, the Peterson fragmentation parameter is set to $\epsilon=0.035$.
 The NLO QCD predictions (figure ~\ref{diffxsections}) are in reasonable agreement with the data, except in the \etads\ distribution, where the measurements show a shift into the positive $\eta$ region (proton direction) with respect to 
the prediction. Also a softer charm fragmentation is 
favored by the data. 
\subsection{Fragmentation Effects}
In Monte Carlo fragmentation  models like JETSET or HERWIG, 
 a forward shift in the \ds\ direction w.r.t. that of the original charm quark is produced during the fragmentation due to the interaction of the 
charm quark with the proton remnant via either strings or soft gluon radiatio

To investigate how this affects the NLO QCD predictions 
we have reweighted a LO Monte Carlo for charm production, RAPGAP (it uses 
JETSET for the fragmentation) in a way such that at the stage of 
the hard interaction it will reproduce exactly the HVQDIS results 
for the $p_t(c)$ and $\eta(c)$ differential cross sections. The predictions from this NLO reweighted RAPGAP Monte Carlo are shown in (figure~\ref{diffxsections}). They provide a better descripion of the data, 
especially in the $\eta(\ds)$ and $x(\ds)$ differential cross sections
  This result suggest that the small disagreement found with HVQDIS come from the fact that the Peterson 
function can not account for all the charm quark fragmentation effects present at HERA, in particular the interaction with the remnant.

\begin{figure}[htb]
\epsfig{file=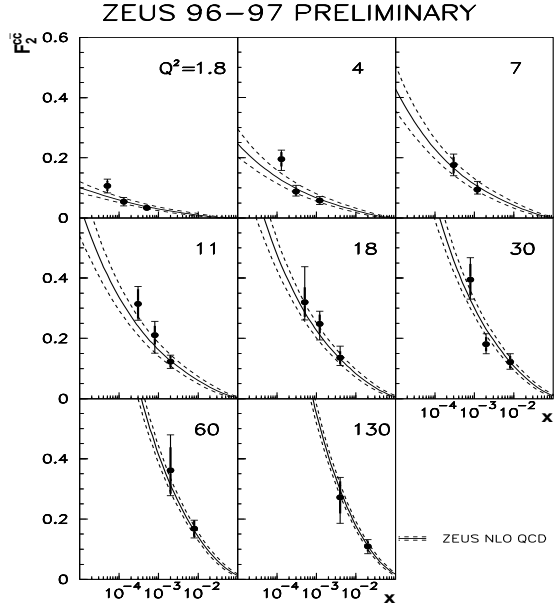,width=11cm,height=8.5cm}
\vspace{-1.7cm}
\caption{ The measured \ftwoccb\ at constant \qsq\ 
as a function of $x$. 
The result of the NLO QCD ZEUS 
fit ~\cite{zeusfit} is the solid curve. The dashed curves show the error from that fit.}
\label{f:f2c}
\end{figure}
\vspace{-0.2cm}
\section{\ftwoccb\ EXTRACTION}
\vspace{-0.1cm}
 The procedure for extracting \ftwoccb\ 
   starts by measuring the \dspm\ production cross section in the 
   restricted \ptds, \etads\ region in bins of \qsq\ and $y$. 
   The extrapolation to the full \ptds, \etads\  
   region is done in the following way:
\vspace{-0.1cm}
\begin{equation}
\ftwoccb^{\mbox{m}}(x_i, Q^2_i) = 
                   \frac{\sigma_i^{\mbox{m}}(e^+p\rightarrow \ds X)}
                        {\sigma_i^{\mbox{t}}(e^+p\rightarrow \ds X)}
                    \ftwoccb^{\mbox{t}}(x_i, Q^2_i)
\label{eq:f2c_extr2}
\end{equation}
where $x_i$, $Q^2_i$ is the center of gravity of the bin $i$ 
 and `m' and `t' denote `measured' and `theoretical' 
respectively.
$\ftwoccb^t$ is taken from ZEUS94~\cite{zeusfit} fit.
For $\sigma_i^t$ we use the reweighted Monte Carlo.
A number of assumptions are implicitly done in this procedure:
\begin{itemize}
\vspace{-0.3cm}
\item The TFNS is valid,
\vspace{-0.3cm}
\item $F^{c\bar{c}}_L$ is negligible($<1\%$ of \ftwo\ in our $Q^2$, $y$ region from calculations based on~\cite{laenen}), 
\vspace{-0.3cm}
\item the value of $f(c\rightarrow\ds)$ measured in $e^+e^-$ is valid also at HERA,
\vspace{-0.3cm}
\item the cross section outside the restricted region is well described by NLO QCD.
\vspace{-0.3cm}
\end{itemize}

Figure ~\ref{f:f2c} shows the measured \ftwoccb\ after combining the results from  both decay channels. Compared to our previous
study we have extended the kinematic range to \qsq\ as low as 1.8~GeV$^2$ and
up to 130~GeV$^2$ and the errors are reduced 
substantialy. \ftwoccb\ exhibits a steep rise
with decreasing $x$ at constant \qsq. From a comparison 
with the ZEUS94 parametrization we determine that \ftwoccb\ 
accounts for $<10$\% of \ftwo\ at low \qsq\ 
and $x\simeq 5\cdot10^{-4}$ and $\simeq 30$\% of \ftwo\ for \qsq$>11$~GeV$^2$ 
at the lowest $x$ measured. 

\section{SUMMARY}
\vspace{-0.1cm}
We have presented a charm analysis in DIS using the combined 1996 and 1997 
data sample. Charm was tagged with \ds\ mesons decaying into two decays ($K2\pi$ and $K4\pi$).
In the experimentally 
accessible region, differential \dspm\ cross sections are in reasonable agreement with NLO QCD 
calculations of charm production in the TFNS using a pdf extracted from an 
     inclusive measurement of \ftwo. This represents a successful test of the universality of the pdf's.

Small disagreements  in the \etads\ and \xd\ distributions show that the
 fragmentation a la Peterson  can not account for all the charm quark fragmentation 
effects present at HERA.

Using these calculations to extrapolate outside the measured
\ptds, \etads\ region, \ftwoccb\ was extracted.
\ftwoccb\ is rising steeply with decreasing $x$ at costant \qsq. It amounts to $\approx 25\%$ of \ftwo\ at low $x$, $\qsq>10$.


\begin{thebibliography}{9}
\vspace{-0.1cm}
\bibitem{h194} H1 Collab., C.~Adloff {\it et al.},\\ Z.~Phys.~C{\bf 72}(1996)593

\bibitem{z94} ZEUS Collab., J.~Breitweg {\it et al.},\\ Phys.~Lett.~B{\bf 407} (1997) 402 

\bibitem{hvqdis} B. W. Harris and J. Smith, hep-ph/9706334

\bibitem{peterson}
C.~Peterson {\it et al.}, Phys.~Rev.~D{\bf 27}, 105 (1983)

\bibitem{smith} B. W. Harris and J. Smith, Nucl. Phys. B{\bf 452}(1995)109;

\bibitem{laenen}
E. Laenen, S. Riemersma, J. Smith and W. L. Van~Neerven, Nucl. Phys. B{\bf 392}(1995)162


\bibitem{zeusfit} ZEUS Collab., J.~Breitweg {\it et al.},
The European Physical Journal C7 (1999) 609-630

\bibitem{OPAL222} OPAL Collab., K.~Ackerstaff {\it et al.}, Eur. Phys.~J~C{\bf 1}(1998)439




\end{thebibliography}
\end{document}